\documentclass[preprint,12pt]{elsarticle}

\usepackage{booktabs}
\usepackage{floatrow}
\usepackage[table,xcdraw]{xcolor}
\usepackage{amsmath}
\usepackage{braket}
\usepackage{graphicx}
\usepackage{dcolumn}
\usepackage{bm}
\usepackage[normalem]{ulem}
\usepackage{color}
\usepackage{hyperref}
\usepackage{multirow}
\usepackage{float}
\usepackage{microtype}
\hypersetup{
colorlinks=true,
citecolor=blue,
linkcolor=blue,
urlcolor=blue
}

\newcommand{\alga}{AlP$_3$/GaP$_3$}
\newcommand{\alpb}{AlP$_3$/PbP$_3$}
\newcommand{\bial}{BiP$_3$/AlP$_3$}
\newcommand{\biga}{BiP$_3$/GaP$_3$}
\newcommand{\gapb}{GaP$_3$/PbP$_3$}
\newcommand{\gesb}{GeP$_3$/SbP$_3$}
\newcommand{\sbbi}{SbP$_3$/BiP$_3$}

\newcommand{\snga}{SnP$_3$/GaP$_3$}
\newcommand{\snpb}{SnP$_3$/PbP$_3$}
\newcommand{\xp}{XP$_3$}
\newcommand{\xaxb}{X$_{\rm{A}}$P$_3$/X$_{\rm{B}}$P$_3$}
\newcommand{\dxab}{$d_{\rm{X_AX_B}}$}

\newcommand{\orcid}[1]{\href{https://orcid.org/#1}{\includegraphics[width=8pt]{orcid.pdf}}}

\begin{document}

\begin{frontmatter}

\title{Descriptor-Based Classification of Interfacial Electronic Coupling in Janus XP$_3$-Based 2D Heterostructures}

\author[ufmt,INCT]{Erika N. Lima}
\ead{erika.lima@fisica.ufmt.br}
\cortext[cor1]{Corresponding author}
\author[ufmt,INCT]{Teldo A. S. Pereira}

\author[ufmt]{Elisangela S. Barboza}

\author[iftm]{Dominike Pacine}

\author[ufla]{Igor S. S. de Oliveira }

\affiliation[ufmt]{
organization={Instituto de Física, Universidade Federal de Mato Grosso},
city={Cuiabá},
postcode={78060-900},
state={MT},
country={Brazil}
}
\affiliation[INCT]{
organization={National Institute of Science and Technology on Materials Informatics},
city={Campinas},
state={SP},
country={Brazil}
}

\affiliation[iftm]{
organization={Instituto Federal de Educação, Ciência e Tecnologia do Triângulo Mineiro},
city={Uberlândia},
postcode={38064-190},
state={MG},
country={Brazil}
}

\affiliation[ufla]{
organization={Departamento de Física,Universidade Federal de Lavras},
city={Lavras},
postcode={37203-202},
state={MG},
country={Brazil}
}

\begin{abstract}

Understanding and controlling interfacial electronic coupling in two-dimensional (2D) heterostructures is essential for designing functional materials for electronic, optoelectronic, and catalytic applications. Here, we investigate vertical heterobilayers constructed from two distinct XP$_3$ monolayers (X = As, Ge, Sb, Bi, Sn, Al, Ga, and Pb) using first-principles density functional theory. The resulting Janus heterobilayers are energetically favorable and elastically stable, with electronic band gaps ranging from metallic and near-metallic to semiconducting regimes. Interlayer interactions induce significant band renormalization, including transitions between type-I and type-II alignment upon structural relaxation. To rationalize these effects, we establish a descriptor-based framework based on the metal-metal interlayer distance, interfacial electron localization, and Bader charge redistribution. This combined analysis discriminates vdW-like, polar–covalent, and ionic interaction regimes, with systematic trends governed by the average atomic number of the constituent elements. Optical absorption calculations indicate visible-to-near-infrared activity in selected systems, and band-edge alignment identifies promising candidates for selective redox processes. Overall, the proposed descriptor-based strategy provides a physically grounded route for identifying and engineering interfacial coupling in XP$_3$ heterostructures and can be extended to other classes of 2D material interfaces.

\end{abstract}


\end{frontmatter}

\section{Introduction}

Two-dimensional (2D) materials provide a versatile platform for engineering interfacial electronic properties through vertical stacking. In such heterostructures, the electronic response is governed by the nature of the interlayer interaction, which controls charge redistribution, band alignment, and carrier separation. Although many 2D systems are described as weakly coupled by van der Waals (vdW) forces, this approximation is not universally valid, as the interaction regime depends on chemical composition and interlayer geometry.~\cite{castellanos2022van}

Among emerging 2D materials, triphosphides with general formula XP$_3$, where X belongs to groups 13–16 and P denotes phosphorus, form a structurally robust and chemically versatile family. Their puckered lattice geometry ensures structural stability, while their electronic properties vary significantly with the identity of X, enabling tunable band gaps, carrier mobility, and optical response.~\cite{GeP3,SbP3-GaP3,SnP3-1} Representative members such as GeP$_3$, SnP$_3$, and SbP$_3$ exhibit narrow band gaps, strong optical absorption, and high carrier mobility. Experimental realization of bulk layered GeP$_3$ and SnP$_3$, as well as few-layer nanosheets, further supports the viability of this materials family.\cite{Donohue1970,Gullman1972,Qi2017-GeP3,Yan2024SnP3}

Recent studies have extended XP$_3$ materials to heterostructure architectures. Oliveira et al. demonstrated that few-layer BiP$_3$ enables Schottky barrier modulation in graphene-based vdW heterostructures.\cite{Bip3-Igor} Lu et al. investigated vertical and lateral XP$_3$ heterostructures, identifying AlP$_3$/GaP$_3$ as a promising photocatalytic system.~\cite{lateral-XP3} 
However, a systematic and physically grounded framework capable of classifying interfacial electronic coupling in heterobilayers composed exclusively of XP$_3$ monolayers is still lacking, as interlayer separation alone is insufficient to classify the interaction regime.

To address this gap, we employ first-principles calculations based on the Density Functional Theory (DFT) to investigate vertical heterobilayers formed by stacking two distinct XP$_3$ monolayers (\xaxb). These heterobilayers exhibit a Janus character as a result of compositional asymmetry between the layers, generating non-equivalent interfacial environments. To ensure realistic and experimentally viable systems, our analysis is restricted to thermally and dynamically stable monolayers, namely AsP$_3$, GeP$_3$, SbP$_3$, BiP$_3$, SnP$_3$, AlP$_3$, GaP$_3$, and PbP$_3$.

Here, we establish a descriptor-based framework that quantitatively links geometric and electronic descriptors to the strength and character of interfacial electronic coupling in \xp\ heterobilayers. The metal–metal interlayer distance serves as the primary geometric descriptor, while interfacial electron localization and Bader charge redistribution capture electronic sharing and charge transfer across the interface. Their combined analysis enables discrimination among vdW-like, polar–covalent, and ionic interaction regimes, with deviations from simple distance-based trends rationalized by the average atomic number of the constituent elements. In addition, the physically interpretable descriptors introduced here may provide suitable input features for future data-driven or machine-learning approaches aimed at the rapid screening of heterobilayers.

In addition to classifying interfacial coupling, we analyze mechanical properties, optical absorption spectra, and band-edge alignment relevant to redox processes. Together, these findings provide a rational design roadmap for constructing XP$_3$ heterobilayers with controlled interfacial interaction regimes and tailored functional properties. The descriptor-based strategy proposed here may be extended to other 2D heterostructures in which interlayer electronic coupling plays a central role, provided that the specific chemical and structural characteristics of each system are properly considered.

\section{Computational Method}

All structural and electronic properties in the present work were calculated using first-principles methods based on Density Functional Theory (DFT),\cite{DFT_1,DFT_2} as implemented in the Vienna \textit{ab initio} Simulation Package (VASP).\cite{VASP} The exchange–correlation potential was described within the generalized gradient approximation (GGA) using the Perdew–Burke–Ernzerhof (PBE) functional. The interactions between valence electrons and ionic cores were treated using the projector augmented wave (PAW) method.\cite{PAW_1,PAW_2} Long-range dispersive interactions were included through the semiempirical DFT-D3 correction scheme proposed by Grimme.\cite{grimme2010consistent} This approach provides an accurate description of dispersive forces governing the interlayer binding in layered heterostructures.

The electronic wave functions were expanded in a plane-wave basis set with an energy cutoff of 500 eV. Brillouin zone integrations were performed using a $10 \times 10 \times 1$ $\Gamma$-centered Monkhorst–Pack \textit{k}-point mesh.\cite{Monkhorst} Structural models were constructed using periodic supercells in the \textit{xy}-plane, with a vacuum region of approximately 15~\AA\ introduced along the \textit{z}-direction to avoid spurious interactions between periodic images. A dipole correction along the out-of-plane direction was applied to eliminate artificial electrostatic interactions arising from the asymmetric slab geometry. Structural relaxations were performed using a conjugate-gradient algorithm until the total energy converged to within $10^{-6}$~eV and the residual forces on each atom were below 0.01 eV/\AA. To minimize strain effects and ensure structural compatibility, the lattice mismatch between the stacked monolayers was constrained to a maximum of 2\%, symmetrically distributed between the two constituent layers during structural optimization.

To obtain more accurate electronic band gaps, the electronic band structures were subsequently recalculated using the screened Heyd–Scuseria–Ernzerhof (HSE06) hybrid functional,~\cite{HSE1,HSE2} including spin–orbit coupling (SOC) effects. In these calculations, 25\% of the short-range Hartree–Fock exact exchange was incorporated into the PBE exchange, employing the standard screening parameter of 0.2~\AA$^{-1}$.

The optical properties of the \xp~ heterobilayers were investigated using the WanTiBEXOS code.~\cite{Dias_108636_2023} The Tight-Binding (TB) Hamiltonian was constructed based on the first-principles electronic structure, obtained via the HSE06 hybrid functional using the interface between the VASP and Wannier90 packages.~\cite{Mostofi2014} Maximally Localized Wannier Functions (MLWFs) were generated by considering the projections of the $s$, $p$, and $d$ orbitals of the XP$_3$ heterobilayers. Regarding the interaction model, a truncated Coulomb potential for 2D systems (V2DT) ~\cite{Rozzi2006} was adopted, with a k-point mesh density of $120$ \AA{}$^{-1}$. For the real and imaginary parts of the dielectric function, a broadening parameter of $0.01$ eV was used. The optical spectra were computed in the energy range from 0 to 4 eV, including the eight highest valence bands (n$_v$) and the eight lowest conduction bands (n$_c$). Finally, the optical properties were analyzed at both the Independent Particle Approximation (IPA) level and within the Bethe–Salpeter Equation (BSE) formalism, ~\cite{Salpeter1951} which explicitly accounts for excitonic effects.

\section{Results and Discussion}

\subsection{Structural and Energetic Properties}

We begin our investigation by constructing the individual XP$_3$ monolayers, where X = As, Ge, Sb, Bi, Sn, Al, Ga, and Pb. All systems adopt a puckered hexagonal honeycomb lattice, consistent with prior theoretical predictions for this class of materials.~\cite{ramzan2019electronic} After relaxation, we extracted the equilibrium in-plane lattice constant ($a$) and the monolayer thickness ($h$), defined as the vertical distance between the outermost atoms. Additionally, we calculated the electronic band gap ($E_g$) of each system. 
The calculated structural parameters and electronic band gaps are summarized in Table~\ref{tab:MLs}. The lattice parameter and monolayer thickness increase with the atomic radius of X, with $a$ ranging from 6.73~\AA{} (AsP$_3$) to 7.28~\AA\ (PbP$_3$), and $h$ from 1.49 to 2.96~\AA\ for the same compounds. The band gaps span a wide range, from 0.59 eV in GeP$_3$ to 2.61 eV in AsP$_3$, highlighting the tunability of the electronic properties across the XP$_3$ series. These results are consistent with previous work.~\cite{gama2025dispersive}

\begin{table}
\caption{\label{tab:MLs}In-plane lattice constant $a$, monolayer thickness $h$, and electronic band gap E$_g$ calculated for dynamically and thermally stable XP$_3$ isolated monolayers.}
\begin{tabular}{cccc}
\hline
Monolayer & $a$ (\AA) & $h$ (\AA) & $E_g$ (eV) \\
\hline
AlP$_3$ & 7.21 & 1.39 & 1.50 \\
GaP$_3$ & 7.21 & 1.22 & 1.51 \\
GeP$_3$ & 6.96 & 2.40 & 0.59 \\
SnP$_3$ & 7.15 & 2.85 & 0.74 \\
PbP$_3$ & 7.28 & 2.96 & 0.86 \\
AsP$_3$ & 6.73 & 1.49 & 2.61  \\
SbP$_3$ & 7.00 & 1.79 & 2.34 \\
BiP$_3$ & 7.13 & 1.94 & 2.05 \\
\hline
\end{tabular}
\end{table}

To construct structurally and energetically viable \xaxb\ heterobilayers, we systematically evaluated the lattice mismatch between all possible pairs of monolayers considered in this study.
Only combinations exhibiting a lattice mismatch below approximately 2.0\% were selected, ensuring mechanical compatibility and minimizing interfacial strain in the resulting heterostructures. To further preserve structural symmetry and avoid the concentration of strain in a single layer, the total lattice mismatch was symmetrically distributed between the two monolayers during structural optimization. Based on these criteria, we selected ten representative \xaxb\ heterostructures for detailed investigation. These systems span a broad range of lattice mismatch values, from virtually strain-free configurations such as AlP$_3$/GaP$_3$ (0.00\%) to moderate mismatch systems like SnP$_3$/PbP$_3$ (1.79\%) and SbP$_3$/BiP$_3$ (1.82\%). The smallest mismatches are found in heterostructures such as GeP$_3$/SbP$_3$ (0.57\%) and SnP$_3$/GaP$_3$ (0.83\%), which are expected to exhibit minimal strain-induced distortions and high interfacial quality. On the other hand, even heterostructures with larger mismatch values remain within acceptable limits for theoretical modeling. The complete list of lattice mismatch values for all investigated \xp~heterostructures is presented in Table~\ref{tab:HTs}. 

\begin{figure*}
    \centering
    \includegraphics[width=\textwidth]{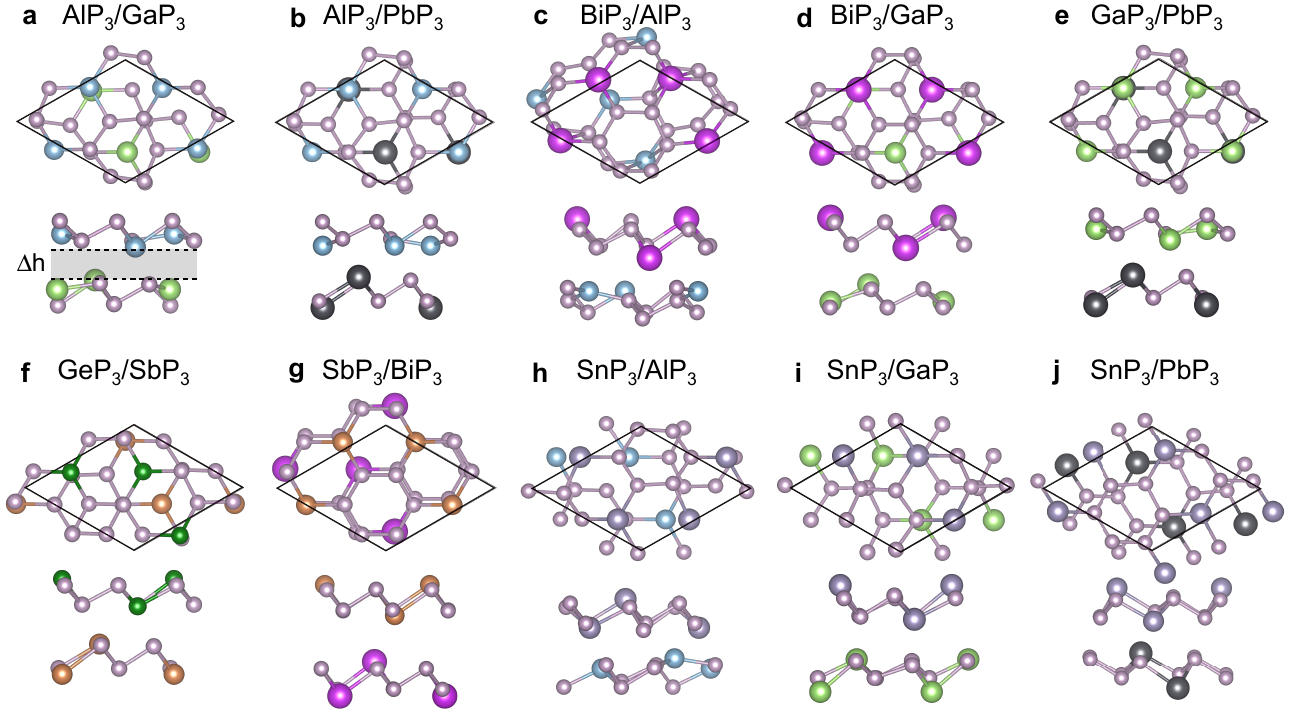}
   \caption{Top and side views of the optimized atomic structures of the investigated Janus XP$_3$ heterobilayers: (a) AlP$_3$/GaP$_3$, (b) AlP$_3$/PbP$_3$, (c) BiP$_3$/AlP$_3$, (d) BiP$_3$/GaP$_3$, (e) GaP$_3$/PbP$_3$, (f) GeP$_3$/SbP$_3$, (g) SbP$_3$/BiP$_3$, (h) SnP$_3$/AlP$_3$, (i) SnP$_3$/GaP$_3$, and (j) SnP$_3$/PbP$_3$. The dashed line represents the interlayer distance $\Delta h$, defined as the vertical separation between the metal atoms located at the centers of the two opposite layers forming the heterobilayer. The atomic species are identified by the following color code: light blue for Al, light green for Ga, dark gray for Pb, magenta for Bi, dark green for Ge, orange for Sb, purple for Sn, and light pink for P.}
    \label{fig:structures}
\end{figure*}

\begin{table*}
\caption{\label{tab:HTs} Properties of representative \xp\ heterostructures: 
lattice mismatch (strain), interlayer spacing 
($\Delta h$), interlayer binding energy ($E_{\rm b}$), electronic band gap ($E_g$), equilibrium interlayer distance between the metals X$_A$ and X$_B$ (\dxab), electron localization function evaluated at the metal--metal midpoint (ELF$_\mathrm{mid}$), and the average atomic number $\bar{Z}$, which is defined as 
$\bar{Z}=(Z_{\rm{X_A}}+Z_{\rm{X_B}})/2$.}
\begin{tabular}{cccccccccc}
\hline
\multirow{2}{*}{Heterobilayers} & strain & $\Delta h$ & $E_{\rm b}$ & $E_g$ & $\Delta\rho$ & \dxab & ELF$_\mathrm{mid}$ & $\bar{Z}$ \\
& (\%) & (\AA) & (meV/\AA$^2$) & (eV) & ($10^{13} e/\rm{cm}^{2}$) & (\AA) & ($-$) & ($-$) \\
\hline
AlP$_3$/GaP$_3$ & 0.00 & 1.85 & -143.5 & 0.35 & 9.61  & 2.94 & 0.30 & 18.0 \\
AlP$_3$/PbP$_3$ & 0.96 & 1.83 & -141.5 & 0.67 & 14.40 & 3.55 & 0.03 & 47.5 \\
BiP$_3$/AlP$_3$ & 1.11 & 1.77 & -121.3 & 0.61 & 5.92  & 3.42 & 0.31 & 48.0 \\
BiP$_3$/GaP$_3$ & 1.11 & 2.01 & -125.3 & -    & 1.53  & 4.59 & 0.02 & 57.0 \\
GaP$_3$/PbP$_3$ & 0.96 & 2.00 & -131.6 & 0.09 & 3.09  & 4.64 & 0.02 & 56.5 \\
GeP$_3$/SbP$_3$ & 0.57 & 2.04 & -140.1 & 0.11 & 3.57  & 2.92 & 0.60 & 41.5 \\
SbP$_3$/BiP$_3$ & 1.82 & 2.66 & -109.7 & 1.11 & 0.46  & 5.82 & 0.23 & 67.0 \\
SnP$_3$/AlP$_3$ & 0.83 & 1.74 & -141.3 & 0.89 & 16.60 & 2.69 & 0.22 & 32.5 \\
SnP$_3$/GaP$_3$ & 0.83 & 2.26 & -136.2 & -    & 1.46  & 4.24 & 0.02 & 41.5 \\
SnP$_3$/PbP$_3$ & 1.79 & 1.89 & -129.6 & -    & 0.90  & 3.28 & 0.23 & 66.0 \\
\hline
\end{tabular}
\end{table*}

The binding energies of the heterobilayers are computed according to 
\begin{equation}
    E_{\rm b} = (E_{\rm het} - E_{\rm X_AP_3} - E_{\rm X_BP_3})/S,
\end{equation}
where $E_{\rm het}$ and $E_{\rm X_{A/B}P_3}$ denote the total energies of the heterostructure and the isolated \xp\ monolayers, respectively, and $S$ is the supercell area.
Several initial stacking configurations between the two monolayers were considered. The results corresponding to the most stable $E_{\rm b}$ values are presented in Table~\ref{tab:HTs}, and the optimized structures are depicted in Fig.~\ref{fig:structures}.
 Table~\ref{tab:HTs} shows that XP$_3$ heterobilayers exhibit strong interlayer binding energies, with adsorption energies ranging from 
–129.6~meV/\AA$^2$ (\sbbi) to –143.5~meV/\AA$^2$ (\alga), a variation of $\sim$10\%. 
The corresponding interlayer spacings vary between 1.74 and 2.66~\AA{}, reflecting subtle differences in atomic size and orbital overlap across the different structures. Heterostructures involving lighter elements (e.g., Al, Ga) tend to exhibit shorter separations, consistent with stronger binding, while those containing heavier elements (e.g., Sb, Bi) show larger spacings and weaker interactions.

\subsection{Electronic Properties}

Next, we investigate the charge transfer ($CT$) between the monolayers in the \xp\ heterostructures. We use the Bader charge analysis \cite{Bader,Henkelman} to compute the 
net charge gain or loss of each XP$_3$ layer, which was obtained by summing the Bader charges of all atoms belonging to a given layer in the heterostructure and subtracting the corresponding charge of the isolated free-standing XP$_3$ monolayer. The differential charge density $\Delta\rho$ was then calculated by normalizing the resulting difference by the in-plane supercell area $S$. 

The results for $\Delta\rho$
presented in Table~\ref{tab:HTs} reveal systematic trends that complement understanding of the variations in interlayer binding energies and interlayer spacings across the \xaxb\ heterostructures. The interfaces involving AlP$_3$ show the largest charge redistribution, with densities reaching
$9.6\times 10^{13}-1.7\times 10^{14}~e/\rm{cm}^2$ when paired with GaP$_3$, PbP$_3$, and SnP$_3$. These values correlate directly with the strongest interlayer binding observed in the series, between -143.5 to -141.3~meV/\AA$^2$, which also display shorter interlayer distances
($\sim 1.7-1.9$~\AA). This indicates that stronger orbital coupling at small separations is supported by modest but significant electron redistribution at the interface.

By contrast, heterostructures composed of combinations of monolayers with heavier elements, such as BiP$_3$, SbP$_3$, SnP$_3$, and PbP$_3$ exhibit much smaller $CT$ densities, such as the BiP$_3$/SbP$_3$ interface ($4.6\times 10^{12}~e/\rm{cm}^2$) and the SnP$_3$/PbP$_3$ interface
($9.0\times 10^{12}~e/\rm{cm}^2$), both of which correspond to larger interlayer spacings and weaker interlayer binding energies. These reduced values reflect the more ideal vdW character of the interaction, where limited orbital overlap leads to minimal charge redistribution.

\begin{figure*}[!htb]
    \centering
    \includegraphics[width=\textwidth]{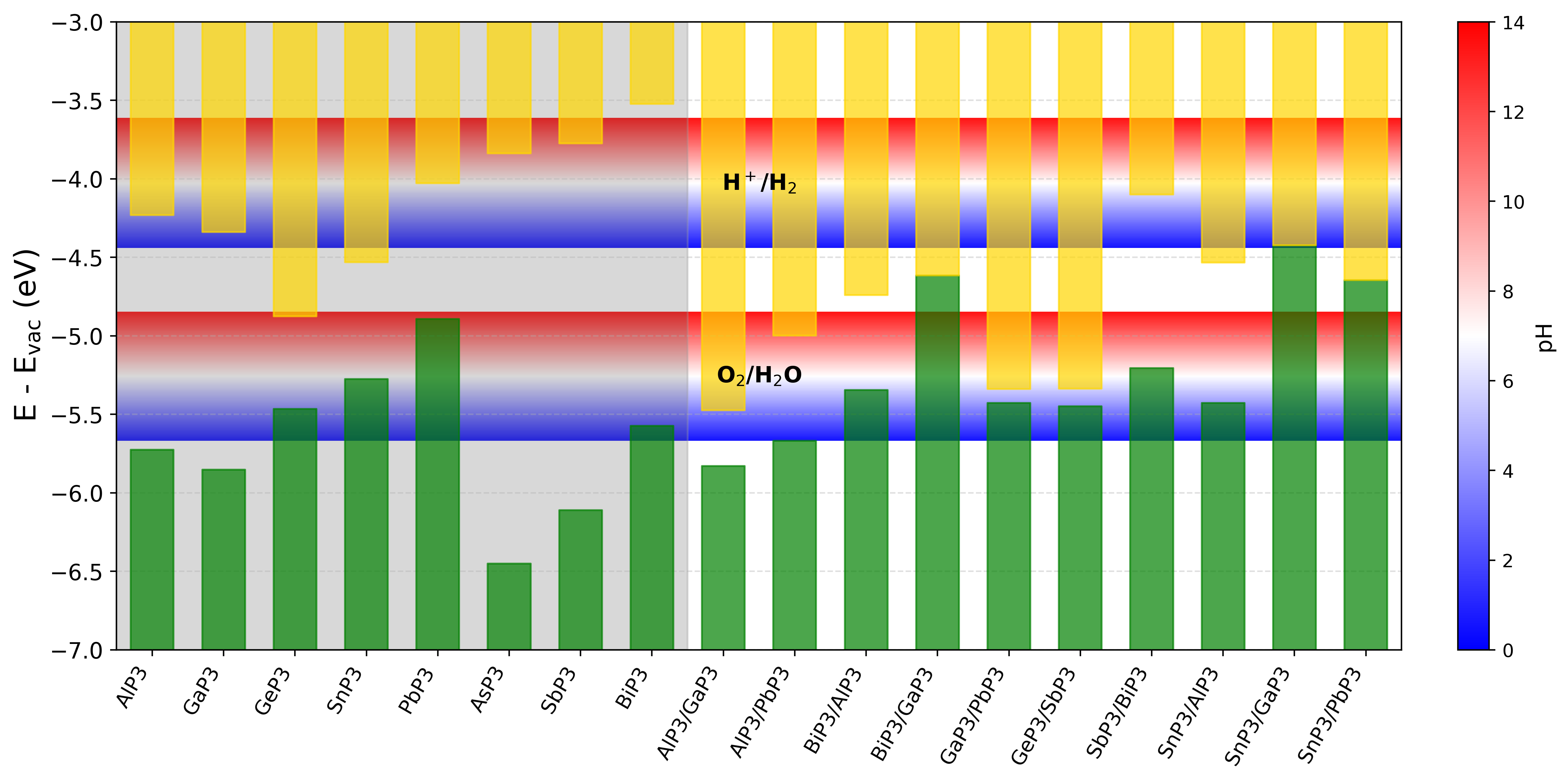}
    \caption{Band edge alignment of the monolayers and heterostructures relative to the vacuum level (E - E$_{\rm{vac}}$). The yellow bars represent the conduction band minimum (E$_{\rm{CBM}}$), and the green bars correspond to the valence band maximum (E$_{\rm{VBM}}$). The horizontal colored regions indicate the redox potentials of water, with the upper band corresponding to the H$^{+}$/H$_{2}$ level and the lower band to the O$_{2}$/H$_{2}$O level. The color map represents the pH dependence of the redox potentials, varying continuously from pH 0 (blue) to pH 14 (red), as described by the Nernst equations \ref{eq:1} and \ref{eq:2}.}
    \label{fig:redox}
\end{figure*}

In Fig.~\ref{fig:redox}, we present the band-edge positions of the \xp\ monolayers together with those of the investigated \xaxb\ heterostructures. In several cases, the direction of charge transfer follows the trend expected from the ionization energy ($IE$) differences, with electrons consistently flowing from layers with lower $IE$ (donors) to those with higher $IE$ (acceptors). This behavior is observed, for example, in heterostructures where GaP$_3$ receives electrons from AlP$_3$, BiP$_3$, or SnP$_3$.
However, in other systems, charge transfer ($CT$) does not follow the trend predicted by the $IE$ differences, such as in the heterostructures formed by AlP$_3$ with PbP$_3$, BiP$_3$, or SnP$_3$, as well as PbP$_3$ combined with GaP$_3$ or SnP$_3$. These deviations indicate that the direction and magnitude of charge transfer
are primarily governed by interfacial phenomena, particularly orbital hybridization, chemical bonding strength, and the formation of interface dipoles.
In general, the observed $CT$ behavior reveals that interfaces exhibiting larger electron redistribution tend to show stronger interlayer binding and shorter equilibrium separations, while systems with minimal charge transfer correspond to weak, predominantly vdW-coupled heterostructures. 
\begin{figure*}
    \centering
    \includegraphics[width=\textwidth]{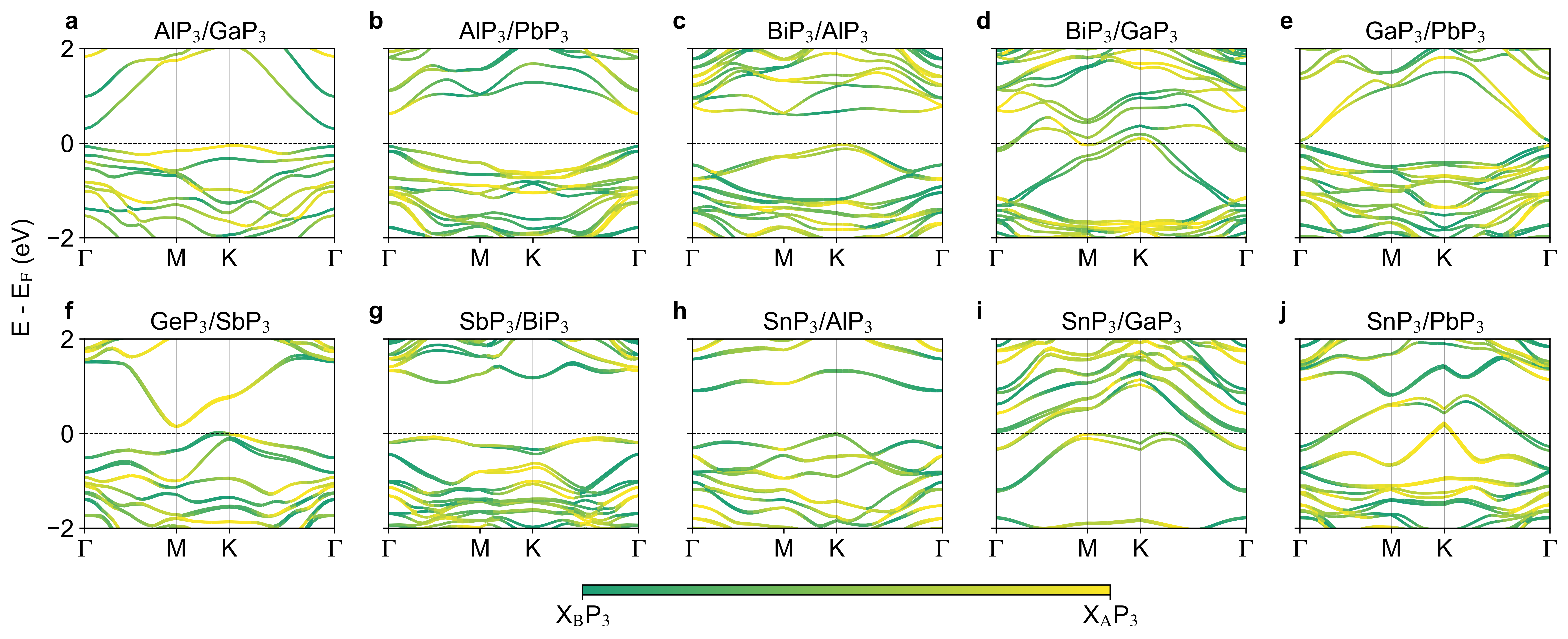}
  \caption{Layer-projected electronic band structures of representative \xp\ heterostructures: (a) AlP$_3$/GaP$_3$, (b) AlP$_3$/PbP$_3$, (c) BiP$_3$/AlP$_3$, (d) BiP$_3$/GaP$_3$, (e) GaP$_3$/PbP$_3$, (f) GeP$_3$/SbP$_3$, (g) SbP$_3$/BiP$_3$, (h) SnP$_3$/AlP$_3$, (i) SnP$_3$/GaP$_3$, and (j) SnP$_3$/PbP$_3$. The dashed horizontal line indicates the Fermi level set to zero energy. The color scale represents the relative contribution of each constituent monolayer to the electronic states, allowing identification of band localization and interlayer hybridization near the valence and conduction band edges.}
    \label{fig:bandproj}
\end{figure*}

The XP$_3$ heterobilayers exhibit a wide range of band gaps, as presented in Table~\ref{tab:HTs}, spanning from metallic (\biga, \snga, \snpb) and nearly-metallic values of 0.09--0.11~eV in \gapb\ and \gesb\ to semiconducting gaps close to 1.1~eV in \sbbi. In general, stacking reduces the band gap relative to the isolated monolayers (0.59--2.61~eV), reflecting enhanced interlayer hybridization. 
The most pronounced narrowing occurs in heterostructures containing heavier elements such as Sn, Pb, and Bi, where strong orbital overlap and SOC favor charge delocalization. Conversely, combinations that pair lighter (Al, Ga, As) with heavier elements retain intermediate gaps of 0.6--0.9~eV, balancing carrier transport with infrared optical activity. Altogether, the results demonstrate that \xp\ heterostructures provide a versatile platform where the electronic band gap can be tuned from a near-metallic regime to the visible–IR range, enabling potential applications in low-power electronics, and optoelectronics.

\subsection{Interlayer Interaction Regimes}

While the previous sections established systematic trends in binding energy and charge transfer, a deeper understanding of the mechanisms governing interlayer coupling is still required. Therefore, we introduce a unified and physically grounded descriptor framework that distinguishes ionic, polar–covalent, and vdW-like interaction regimes and enables predictive screening of \xaxb\ heterostructures based on their intrinsic atomic characteristics.

In this framework, the equilibrium metal–metal interlayer distance \dxab, the electron localization function (ELF) at the interface, and the Bader charge redistribution $\Delta\rho$ are employed as complementary descriptors of the interlayer interaction. Together, these quantities capture the geometric configuration of the interface, the degree of electronic localization, and the magnitude of interlayer charge transfer, providing a comprehensive description of the coupling mechanisms between stacked layers.

\begin{figure*}
    \centering
    \includegraphics[width=\textwidth]{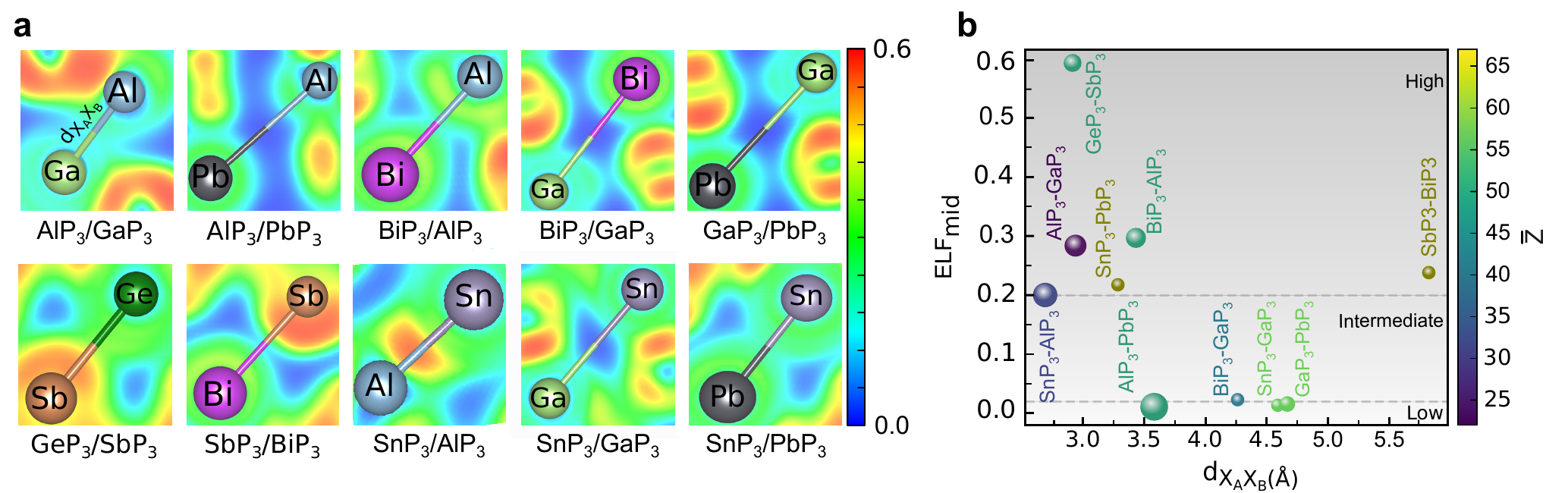}
    \caption{(a) Electron localization function (ELF) mapped along the interlayer direction across the interface, illustrating the spatial evolution of electronic localization between the stacked XP$_3$ monolayers. (b) Scatter plot of ELF$_\mathrm{mid}$ as a function of the equilibrium interlayer distance \dxab, where the marker size is proportional to the magnitude of the Bader charge transfer $|\Delta \rho|$, and the color scale represents the average atomic number $\bar{Z}$ of the constituent metal atoms.}
    \label{fig:ELF_IC}
\end{figure*}

Within this framework, we analyze the interlayer interaction through the electron localization function evaluated at the midpoint of the interlayer metal–metal contact, denoted as ELF$_\mathrm{mid}$. This quantity corresponds to the ELF value at the geometric midpoint of the equilibrium metal–metal interlayer distance \dxab, which defines the central region of the interface, as illustrated in Fig.~\ref{fig:ELF_IC}(a). The use of ELF$_\mathrm{mid}$ is motivated by its direct sensitivity to electronic localization in the interfacial region where interlayer coupling is established. Unlike the Bader charge transfer $\Delta\rho$, which quantifies the net electron redistribution between layers, ELF$_\mathrm{mid}$ probes the degree of electronic sharing or depletion in real space, providing insight into whether the interaction is dominated by covalent overlap, ionic polarization, or weak dispersive forces.

By evaluating ELF specifically at the metal–metal midpoint, we focus on the region where orbital overlap and interfacial hybridization are maximized, making ELF$_\mathrm{mid}$ a direct indicator of the strength and character of interlayer electronic coupling. When combined with the Bader charge transfer $\Delta\rho$, which captures long-range electrostatic redistribution, this approach enables a unified and physically transparent classification of interaction regimes.
In addition to these primary descriptors, the average atomic number $\bar{Z}$ is considered as an auxiliary parameter to provide physical insight into deviations from the general ELF$_\mathrm{mid}$--$\Delta\rho$--\dxab\ trends. The obtained quantity values are summarized in Table~\ref{tab:HTs}.

To visualize the relationship between electronic localization and interlayer separation, Fig.~\ref{fig:ELF_IC}(b) presents ELF$_\mathrm{mid}$ as a function of the equilibrium metal–metal interlayer distance \dxab. The size of the markers represents the magnitude of the Bader charge transfer $|\Delta\rho|$, while the color scale indicates the average atomic number $\bar{Z}$ of the constituent elements.

Both ELF$_\mathrm{mid}$ and $\Delta\rho$ exhibit a clear inverse dependence on the equilibrium interlayer distance \dxab. High ELF$_\mathrm{mid}$ values ($\sim$0.20–0.60), together with substantial charge transfer, occur at short interlayer distances (\dxab $< 3.0$~\AA), as observed for AlP$_3$/GaP$_3$ and SnP$_3$/AlP$_3$, characterizing ionic or polar–ionic coupling. GeP$_3$/SbP$_3$ also lies in this short-distance regime but exhibits moderate charge redistribution combined with high ELF$_\mathrm{mid}$, indicating a polar–covalent interaction dominated by electronic sharing. BiP$_3$/AlP$_3$, at an intermediate interlayer distance (3.42~\AA), maintains a relatively high ELF$_\mathrm{mid}$ (0.31) together with moderate charge transfer, placing it in an intermediate polar–covalent regime.
At larger separations (\dxab $> 3.5$~\AA), most interfaces display both low ELF$_\mathrm{mid}$ ($<0.05$) and small charge transfer, as in SnP$_3$/GaP$_3$, BiP$_3$/GaP$_3$, and GaP$_3$/PbP$_3$, consistent with vdW-like or weakly interacting interfaces. AlP$_3$/PbP$_3$ constitutes a clear exception, exhibiting very low ELF$_\mathrm{mid}$ despite a large charge transfer, which indicates a predominantly ionic interaction stabilized by long-range electrostatics rather than electronic overlap.

Notably, SnP$_3$/PbP$_3$ and SbP$_3$/BiP$_3$ stand out as distinctive cases in the high-$\bar{Z}$ regime. Both interfaces are classified as vdW-like, albeit for different physical reasons: SnP$_3$/PbP$_3$ shows negligible charge transfer even at short distances, whereas SbP$_3$/BiP$_3$ retains a finite ELF$_\mathrm{mid}$ at large separations due to strong electronic polarization. The close agreement between the band structures calculated with and without SOC, as shown in Figs.~S1~(g) and (j) of the Supplementary Material, indicates that these features arise primarily from orbital extension effects rather than from SOC.

Overall, the analysis demonstrates that both ELF$_\mathrm{mid}$ and $\Delta\rho$ are primarily governed by the equilibrium interlayer distance \dxab, although their dependence on this distance is not strictly monotonic. This establishes \dxab\ as a natural first descriptor for initiating the search for interfaces with targeted interlayer coupling. Systematic deviations from the general ELF$_\mathrm{mid}$--$\Delta\rho$--\dxab\ trends are consistently explained by the average atomic number $\bar{Z}$, which captures electronic polarization and orbital extension effects beyond purely geometric considerations.

Taken together, these results define a simple and physically grounded screening framework for \xaxb\ heterostructures. By selecting the constituent atoms $X_{\rm{A}}$ and X$_{\rm{B}}$ according to their atomic size and $\bar{Z}$, it is possible to preselect ionic, polar--covalent, or vdW-like coupling regimes at an early design stage. This provides a property-oriented basis for engineering interlayer interactions and guides the targeted design of XP$_3$-based vertical heterostructures.

\subsection{Elastic Properties}

Having classified the \xp\ heterostructures according to their interfacial interaction regimes, we now examine how these regimes influence their elastic response. The calculated elastic constants and derived mechanical parameters are reported in Table~\ref{tab:constants}.
The mechanical behavior reflects the combined influence of interfacial bonding character and the equilibrium metal–metal separation. In all investigated systems, this separation varies from 2.69 to 5.82~\AA{}, establishing a structural scale associated with the degree of interlayer coupling and its impact on in-plane stiffness.

\begin{table}
\caption{\label{tab:constants} Calculated elastic constants (C$_{11}$ and C$_{12}$) and derived mechanical properties, including Young’s modulus (E), shear modulus (G), and Poisson’s ratio ($\nu$), for \xaxb~ heterobilayers. Elastic constants and moduli are expressed in N/m, and Poisson’s ratio is dimensionless.}

\begin{tabular}{cccccccccccc}
\hline
Heterobilayers &  C$_{11}$  & C$_{12}$ & E & G & $\nu$   \\
\hline 
AlP$_3$/GaP$_3$ & 93.05 & 51.59 & 64.44 & 20.73 & 0.55  \\ 
AlP$_3$/PbP$_3$ & 55.53 &49.11 &12.09  &3.21 & 0.88   \\
BiP$_3$/AlP$_3$ & 92.50 &19.72 & 88.29&36.39&0.21\\
BiP$_3$/GaP$_3$ & 90.27 &26.58 & 82.44&31.84&0.29\\
GaP$_3$/PbP$_3$ &87.20  &27.20 & 78.72&30.00&0.31\\ 
GeP$_3$/SbP$_3$ & 127.24 &17.45 & 124.85&54.90&0.14\\ 
SbP$_3$/BiP$_3$ & 98.03 &14.16 & 95.99&41.93&0.14\\ 
SnP$_3$/AlP$_3$ & 89.50 &31.02 & 78.75&29.24&0.35\\
SnP$_3$/GaP$_3$ & 87.93 &17.24 &84.55&35.35&0.20\\
SnP$_3$/PbP$_3$ & 74.80 &16.49 & 71.16&29.15&0.22\\ 
\hline
\end{tabular}
\end{table}

Heterobilayers exhibiting ionic or polar–covalent interaction character, including SnP$_3$/AlP$_3$ (2.69 \AA{}), GeP$_3$/SbP$_3$ (2.92 \AA{}), AlP$_3$/GaP$_3$ (2.94 \AA{}), BiP$_3$/AlP$_3$ (3.42 \AA{}), and AlP$_3$/PbP$3$ (3.55 \AA{}), are characterized by comparatively shorter separations. Reduced distances favor enhanced orbital overlap and electrostatic interaction, which are generally associated with increased in-plane stiffness. Within this group, C${11}$ ranges from 55.53 to 127.24 N/m, Young’s modulus from 12.09 to 124.85 N/m, and the shear modulus from 3.21 to 54.90 N/m. With the exception of AlP$_3$/PbP$_3$, which exhibits comparatively low rigidity, the remaining systems display higher elastic constants consistent with stronger interfacial coupling.

In contrast, heterobilayers classified as predominantly vdW-like, such as SnP$_3$/GaP$_3$ (4.24 \AA{}), BiP$_3$/GaP$_3$ (4.59 \AA{}), GaP$_3$/PbP$_3$ (4.64 \AA), SnP$_3$/PbP$_3$ (3.28 \AA), and SbP$_3$/BiP$3$ (5.82 \AA{}), generally exhibit larger separations and moderate elastic constants. In this regime, C$_{11}$ varies between 74.79 and 98.03 N/m, Young’s modulus between 71.16 and 95.99 N/m, and the shear modulus between 29.15 and 41.93 N/m. The predominance of dispersive interactions in these systems is consistent with comparatively weaker interfacial mechanical coupling.

The partial overlap between the separation intervals (3.28–3.55 \AA{}) indicates that metal–metal distance alone does not fully determine the elastic response. Within this intermediate range, electronic polarization, charge redistribution, and bonding directionality contribute to the observed variations in stiffness. For example, SnP$_3$/PbP$_3$ (3.28 \AA{}) behaves as a vdW-like system despite its relatively short separation, whereas BiP$_3$/AlP$_3$ (3.42 \AA{}) exhibits higher rigidity associated with stronger polar–covalent contributions. Similarly, AlP$_3$/PbP$_3$ (3.55 \AA{}) deviates from a simple distance–stiffness correlation, presenting low Young’s modulus and high Poisson’s ratio, indicative of pronounced transverse strain coupling.

Overall, the equilibrium metal–metal separation serves as a primary structural descriptor correlated with interfacial stiffness, while the bonding regime and electronic structure effects modulate the mechanical response, particularly within the intermediate distance range. These results reinforce that geometric proximity and interfacial electronic character must be considered jointly when analyzing the elastic behavior of \xaxb\ heterostructures.

\subsection{Optical Properties}

The optical response of XP$_3$ heterobilayers provides important insight into their potential for optoelectronic and photocatalytic applications. In 2D systems, reduced dielectric screening and interlayer electronic coupling can significantly modify light–matter interactions and enhance excitonic effects. To characterize the optical behavior of the \xp\ heterostructures under electromagnetic radiation, we analyze the absorption coefficient.

\begin{figure}
    \centering
    \includegraphics[width=\textwidth]{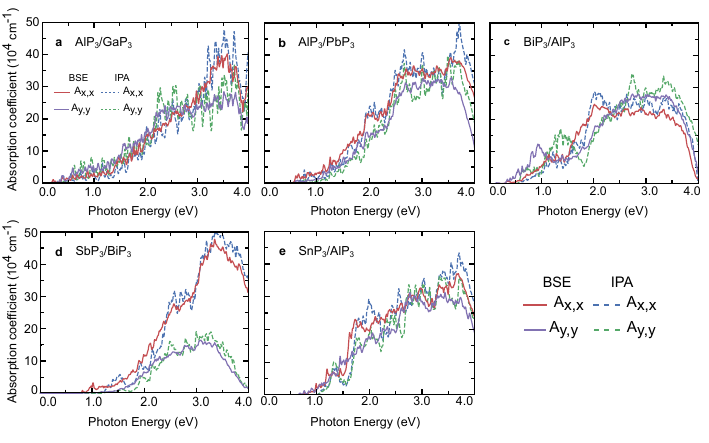}
    \caption{Absorption spectra of representative \xp\ heterostructures: (a) AlP$_3$/GaP$_3$, (b) AlP$_3$/PbP$_3$, (c) BiP$_3$/AlP$_3$, (d) SbP$_3$/BiP$_3$, and (e) SnP$_3$/AlP$_3$. Solid curves correspond to spectra calculated within the Bethe–Salpeter Equation (BSE) formalism, while dashed curves represent the Independent Particle Approximation (IPA). Red and purple lines denote light polarized along the $xx$ and $yy$ directions.}
    \label{fig:optics}
\end{figure}

To illustrate the main optical trends, we focus on semiconducting heterobilayers with moderate band gaps, namely AlP$_3$/GaP$_3$ (0.35 eV), AlP$_3$/PbP$_3$ (0.67 eV), BiP$_3$/AlP$_3$ (0.61 eV), SbP$_3$/BiP$_3$ (1.11 eV), and SnP$_3$/AlP$_3$ (0.89 eV). Fig.~\ref{fig:optics} presents the absorption coefficient for the linear polarization directions $xx$ and $yy$. The results reveal that these compounds are intrinsically anisotropic, exhibiting distinct optical responses depending on the polarization direction of the incident electric field. Furthermore, the \xp\ heterostructures begin to absorb in the infrared region and display pronounced peaks spanning the visible and ultraviolet energy ranges. 
Among the five heterobilayers studied, only the \bial\ system exhibits a more pronounced absorption coefficient for light polarized along the $yy$ direction. This behavior can be attributed to the orbital character of the states forming the valence and conduction band edges. When the contributing orbitals ($s$, $p$, and $d$) are preferentially oriented along a given crystallographic direction, the transition probability for photons polarized along that direction increases. In addition, structural anisotropy arising from lattice distortions or stacking configuration leads to an inequivalent electronic response along the $x$ and $y$ directions.\cite{castellanos2022van}

Notably, AlP$_3$/GaP$_3$, AlP$_3$/PbP$_3$, BiP$_3$/AlP$_3$, and SbP$_3$/BiP$_3$ are found to exhibit type-II band alignment (see Fig.~\ref{fig:redox}). In these systems, the absorption spectra show relatively weak excitonic features near the band gap, characterized by discrete peaks and low absorption intensity. This behavior may be associated with the partial spatial separation of electrons and holes between adjacent layers. Similar effects have been reported in type-II heterostructures such as MoSe$_2$–WSe$_2$.~\cite{Rivera2015} Although a detailed analysis of the excitonic states is beyond the scope of the present work, this charge separation mechanism may be beneficial for photocatalytic applications.

\subsection{Band-Edge Alignment for Photocatalysis}

The photocatalytic performance of the Janus XP$_3$ heterostructures was evaluated through band edge alignment relative to the vacuum level, as shown in Fig.~\ref{fig:redox}. The positions of the conduction band minimum ($E_{\rm{CBM}}$) and valence band maximum ($E_{\rm{VBM}}$) were compared with the pH-dependent redox potentials of water on the absolute energy scale. This approach allows assessment of the thermodynamic feasibility of the hydrogen evolution reaction (HER) and oxygen evolution reaction (OER).

The pH dependence of the redox potentials is described by the Nernst equations:~\cite{nernst-equation}
\begin{equation}
E_{\mathrm{HER}}(pH) = -4.44 + 0.059\,pH \quad \text{(eV)},\label{eq:1}
\end{equation}
\begin{equation}
E_{\mathrm{OER}}(pH) = -5.67 + 0.059\,pH \quad \text{(eV)}.\label{eq:2}
\end{equation}
For photocatalytic water splitting, the band edges of the semiconductor must straddle the redox potentials of water. The thermodynamic criteria for overall water splitting under single-photon excitation can be expressed as:
\begin{equation}
    E_{\rm{CBM}} > E_{\mathrm{HER}}(pH),
\end{equation}
\begin{equation}
    E_{\rm{VBM}} < E_{\mathrm{OER}}(pH),
\end{equation}
\begin{equation}
    E_g \ge 1.23 \ \text{eV},
\end{equation}
where 1.23 eV corresponds to the thermodynamic potential difference between HER and OER at standard conditions. At pH 0, 7, and 14, the redox levels shift according to the above relations, modifying the alignment conditions accordingly.

Fig.~\ref{fig:redox} indicates that seven heterobilayers exhibit semiconducting behavior, while three systems are metallic. Band alignment inferred from the isolated monolayers suggests two type-I systems, GeP$_3$/SbP$_3$ and SnP$_3$/AlP$_3$, and five type-II systems: AlP$_3$/GaP$_3$, AlP$_3$/PbP$_3$, BiP$_3$/AlP$_3$, GaP$_3$/PbP$_3$, and SbP$_3$/BiP$_3$. However, explicit calculations of the complete heterobilayers, as shown in Fig.~\ref{fig:bandproj}, reveal band renormalization induced by interlayer interaction. In particular, GeP$_3$/SbP$_3$ transitions from a type-I alignment predicted for the isolated monolayers to a type-II configuration in the fully relaxed heterobilayer, as shown in Fig.~\ref{fig:bandproj}(f).

According to Eqs.~\ref{eq:1} and \ref{eq:2}, only SbP$_3$/BiP$_3$ exhibits a conduction band minimum (CBM) position compatible with the hydrogen evolution reaction (HER) within the investigated pH range, satisfying the HER criterion only under alkaline conditions. In contrast, all heterostructures meet the thermodynamic requirement for the oxygen evolution reaction (OER) throughout the considered pH window. Although some systems satisfy the band-edge alignment conditions for one or both half-reactions, their relatively small band gaps limit their suitability for spontaneous overall water splitting under single-photon excitation.~\cite{water-spliting,water-spliting-2}

Within this context, among the studied \xp\ heterobilayers, \alpb, \bial, and \sbbi\ stand out as the most promising candidates for selective half-reactions, as their band alignment supports either hydrogen evolution or oxygen evolution depending on the pH conditions. These systems combine semiconducting character, moderate band gaps, and type-II band alignment, which facilitates spatial separation of photogenerated charge carriers and is advantageous for interfacial redox processes.~\cite{photo, photo-2}

To further quantify the driving force behind this spatial separation, we evaluate the intrinsic internal field (E$_{\text{built-in}}$) within each junction using the electrostatic gradient formalism, E$_{\text{built-in}} = -\Delta V/H_{\text{total}}$.~\cite{efield0} Here, $\Delta V$ represents the potential drop between the topmost and bottommost vertical positions of the heterobilayer, and H$_{\text{total}}$ denotes the total thickness, including the interlayer distance $\Delta h$. Notably, \alpb\ and \bial\ exhibit the most robust internal fields in the series, reaching magnitudes of $-4.09 \times 10^7$~V/m and $5.34 \times 10^7$~V/m, respectively. These values are of the same order of magnitude as those reported for high-performance 2D photocatalytic interfaces where strong internal fields are known to effectively suppress radiative recombination.~\cite{efield1, efield2, efield3} The calculated structural parameters, potential drops, and E$_{\text{built-in}}$ magnitudes for all investigated structures are compiled in the Table S1 of the Supplementary Material.

Materials that predominantly facilitate only one half-reaction may nevertheless remain relevant for practical implementations, for instance in the presence of sacrificial electron donors or within tandem and Z-scheme photocatalytic systems, where spatial separation of redox processes can improve the overall efficiency.~\cite{costantino2021sacrificial,schneider2013undesired,pellegrin2017sacrificial,tandem-1,tandem-2}The present assessment focuses exclusively on thermodynamic band alignment, without considering kinetic limitations or reaction overpotentials.

\section{Conclusion}

In summary, we performed a first-principles investigation of vertical heterobilayers formed by \xp\ monolayers. The analyzed systems are energetically favorable and mechanically stable, confirming their viability as 2D heterostructures. The heterobilayers exhibit diverse electronic responses ranging from metallic and near-metallic to semiconducting regimes, with several systems displaying type-II band alignment that favors spatial separation of photogenerated carriers. Selected structures also present visible-to-near-infrared optical absorption and band-edge positions compatible with specific redox half-reactions, indicating potential for optoelectronic and photocatalytic applications. 

To rationalize the interfacial interactions in these systems, we establish a physically grounded framework based on geometric and electronic descriptors that enables the classification and screening of coupling regimes as vdW-like, polar–covalent, or ionic in XP$_3$ heterostructures. Overall, this work establishes a physically grounded strategy for understanding and screening interfacial coupling regimes in XP$_3$-based heterostructures. With appropriate adaptations, the descriptor-based approach introduced here can be extended to other classes of 2D interfaces, offering a general framework for the rational design and discovery of functional 2D heterostructures.

\section*{Acknowledgments}

The authors acknowledge financial support from CNPq, FAPEMIG (including Rede-2D, grant RED-00079-23), FAPEMAT (PRO-2025/00376), INCT Materials Informatics, and INCT de Fotônica. T.A.S.P. and E.S.B. also acknowledge support from PDPG-FAPDF-CAPES Centro-Oeste (00193-00000867/2024-94). The authors further acknowledge the computational resources provided by LCC-UFLA, CENAPAD-SP, and LMCC-UFMT.


\bibliographystyle{elsarticle-num}
\bibliography{reference}

\end{document}